\documentclass[running heads]{llncs}

\usepackage[export]{adjustbox}
\usepackage{booktabs}
\usepackage{ltablex} 
\usepackage{enumitem}
\usepackage{array} 
\usepackage[table]{xcolor}
\usepackage{tabularx}      
\usepackage{ltablex}       
\usepackage{booktabs}      
\usepackage{hyperref}
\usepackage{graphicx}
\usepackage{multicol}
\usepackage{multirow}
\usepackage{siunitx}
\usepackage{subfig}
\usepackage{svg}
\usepackage{tabularx}
\usepackage{textcomp}
\usepackage{xcolor}
\usepackage{xfp}
\usepackage{xspace}
\usepackage{listings}
\usepackage{algorithm}
\usepackage{algpseudocode}

\lstdefinelanguage{yaml}{
  keywords={true,false,null,y,n},
  keywordstyle=\color{darkgray}\bfseries,
  basicstyle=\ttfamily,
  sensitive=false,
  comment=[l]{\#},
  morecomment=[s]{/*}{*/},
  commentstyle=\color{purple}\ttfamily,
  stringstyle=\color{red}\ttfamily,
  morestring=[b]',
  morestring=[b]",
  morestring=[s]{>}{<},
  morestring=[s]{|}{<},
  morestring=[b]` 
}

\definecolor{href:citecolor}{rgb}{.002,.002,.56}
\definecolor{href:linkcolor}{rgb}{.002,.002,.56}
\hypersetup{%
    breaklinks=true,%
    citecolor=href:citecolor,%
    colorlinks=true,%
    linkcolor=href:linkcolor,%
    plainpages=false,%
    urlcolor=href:linkcolor,%
}

\newcolumntype{L}[1]{>{\hsize=#1\hsize\raggedright\arraybackslash}X}%
\newcolumntype{R}[1]{>{\hsize=#1\hsize\raggedleft\arraybackslash}X}%
\newcolumntype{C}[1]{>{\hsize=#1\hsize\centering\arraybackslash}X}%

\expandafter\let\csname c@tblerows\endcsname\rownum

\newcommand{\PCNT}[1]{{\unit[#1]{\kern-1pt\%}}\xspace}

\usepackage[backend=biber, style=ieee]{biblatex}
\begin{document}
\title{Assessing and Prioritizing Ransomware Risk Based on Historical Victim Data}

\titlerunning{Assessing and Prioritizing Ransomware Risk}
\author{Spencer Massengale\inst{1}\orcidID{0009-0003-3483-1697} \and
Philip Huff\inst{1}\orcidID{0000-0003-0869-2147}}


\institute{University of Arkansas at Little Rock, Little Rock, AR 72114, USA}


\vspace{3mm}

\maketitle




\begin{abstract}
We present an approach to identifying which ransomware adversaries are most likely to target specific entities, thereby assisting these entities in formulating better protection strategies. Ransomware poses a formidable cybersecurity threat characterized by profit-driven motives, a complex underlying economy supporting criminal syndicates, and the overt nature of its attacks. This type of malware has consistently ranked among the most prevalent, with a rapid escalation in activity observed. Recent estimates indicate that approximately two-thirds of organizations experienced ransomware attacks in 2023 \cite{Sophos2023Ransomware}. A central tactic in ransomware campaigns is publicizing attacks to coerce victims into paying ransoms. Our study utilizes public disclosures from ransomware victims to predict the likelihood of an entity being targeted by a specific ransomware variant. We employ a Large Language Model (LLM) architecture that uses a unique chain-of-thought, multi-shot prompt methodology to define adversary SKRAM (Skills, Knowledge, Resources, Authorities, and Motivation) profiles from ransomware bulletins, threat reports, and news items. This analysis is enriched with publicly available victim data and is further enhanced by a heuristic for generating synthetic data that reflects victim profiles. Our work culminates in the development of a machine learning model that assists organizations in prioritizing ransomware threats and formulating defenses based on the tactics, techniques, and procedures (TTP) of the most likely attackers.

\keywords{cybersecurity \and LLM \and ransomware.}
\end{abstract}

\section{Introduction}

With a steady rise in the frequency and sophistication of ransomware attacks, organizations are compelled to intensify their efforts in fortifying digital defenses against evolving threats. Ransomware, a form of malware that restricts user access to systems until a ransom is paid, has not only become more prevalent but also increasingly damaging. Organizations of all sizes are affected, with smaller to medium-sized entities disproportionately targeted \cite{Beaman:2021, Richardson_undated-vx}. The consequences of successful ransomware attacks are severe, leading to significant downtime, financial losses, and reputational damage. On average, affected organizations face ransom payouts nearing \$812,000, reflecting the escalating financial toll of this cybersecurity threat \cite{Beaman:2021, Sophos:2022}. Alarmingly, compliance with ransom demands does not always guarantee data recovery, highlighting the complex and evolving challenges posed by ransomware \cite{Sherborne_undated-ls}.

In response to these growing threats, the strategic implementation of Cyber Threat Intelligence (CTI) becomes imperative. Using CTI to understand the adversary can equip entities to proactively defend themselves and improve their ability to make risk-informed decisions \cite{Wagner:2019}. CTI is systematically categorized into tactical, operational, and strategic domains. The tactical facet of CTI focuses on specific Indicators of Compromise (IOCs), such as IP addresses and malware hashes, that facilitate immediate threat detection. Operational CTI provides insights into the Tactics, Techniques, and Procedures (TTPs) employed by adversaries, while strategic CTI delivers an extensive analysis of overarching threat trends, guiding organizational leaders in decision-making and strategic planning \cite{Ettinger2019}.

However, CTI is predominantly consumed through unstructured, natural-language sources such as media reports, news, and bulletins \cite{SANS:2022}, and the effectiveness of CTI is impeded by the overwhelming volume of information, especially for smaller to medium-sized entities \cite{McHugh:2021}. These organizations work with limited resources, hindering their ability to effectively consume CTI \cite{Wagner:2019}. Although these sources offer rich context and real-time insights, assimilating the information into meaningful defensive actions remains challenging. Manual processing becomes impractical with rising numbers of automated systems, sophistication of threats, and growing complexity of data, underlining the urgency for advancements in CTI automation

In this paper, we concentrate on enabling organizations to distill the extensive number of global ransomware threats into a concise list of potential ransomware groups through the application of machine learning techniques. We utilize high-quality data on ransomware variants that are amenable to machine learning analysis and elaborate on our method for extracting pertinent data features through LLMs. Furthermore, we elaborate on our strategy to augment the dataset, ensuring comprehensive coverage of the problem space while preserving attributes that accurately reflect real victim scenarios. Our model uses openly accessible real victim data for analysis, but we recognize the limitations posed by the current scarcity of such data for predicting attacks solely based on existing incidents. To address this gap, we propose a heuristic approach for generating synthetic data to serve as a bridge until the quantity and availability of open-source data are improved. One notable limitation of our method is the temporal evolution of ransomware variants and their target victim profiles, which we address by incorporating time-sensitive characteristics of attacks into our machine learning model. Nevertheless, continuous observation is essential to understand the shifting patterns of ransomware attacks and to update the model's training data accordingly. This methodology is designed to be broadly applicable across various organizational contexts, given that the data features we focus on are universally relevant. The following sections will detail the existing literature, our data management techniques, and provide an in-depth analysis of the machine learning model and its outcomes.

The remainder of this paper is organized as follows: Section \ref{sec:relatedworks} provides a review of related works, highlighting key advancements and methodologies in ransomware threat analysis and predictive modeling. Section \ref{sec:datamanagement} details our methodology for dataset construction and management, describing the criteria for data selection and the techniques employed for feature extraction using LLMs. Section \ref{synth-date-generation} details the process of synthesizing the dataset. In Section \ref{sec:analysis}, we present our approach to ransomware risk analysis, including the development and validation of our predictive model. Section \ref{sec:futurework} discusses the results and implications of our study, emphasizing the effectiveness of the model and its potential application in real-world scenarios. Finally, Section \ref{sec:conclusion} concludes the paper and outlines the predictive capabilities and applicability of our approach in the evolving landscape of cybersecurity threats.

\section{Related Works}
\label{sec:relatedworks}
\subsection{Automating Cyber Threat Intelligence Extraction}
Recent studies have showcased advancements in Cyber Threat Intelligence (CTI) extraction, particularly through the use of Large Language Models (LLMs). Various methods have been developed to refine the automation of extracting different types of CTI, leveraging LLM capabilities for efficient analysis.

The "Time for aCTIon" framework employs GPT-3.5 to streamline CTI extraction using prompt engineering, zero-shot learning, and in-context learning. This approach obviates the need for in-house model training by utilizing LLMs as a Service (LLMaaS), effectively handling diverse CTI scenarios. The framework has been rigorously evaluated against other tools, demonstrating its superiority in quickly and accurately extracting critical CTI elements while optimizing resource utilization \cite{aCTIon}.

Other initiatives, such as TTPDrill and ThreatKG, utilize NLP and information retrieval models to transform adversarial TTPs into structured formats like the Structured Threat Information eXpression  v2.1 (STIX) framework and develop comprehensive threat actor profiles through knowledge graphs \cite{TTPDrill, ThreatKG}. The Ladder framework further extends this by employing models like XLM-RoBERTa and Roberta to associate extracted attack patterns with Mitre Att\&ck IDs, providing a systematic strategy to preempt threats \cite{LADDER, mitre2023attack}.

Further, models like those showcased in CTI View and research by Irshad and Siddiqui extract a combination of strategic CTI and TTPs to pinpoint specific threat actors behind cyber-attacks, aiding the analysis of APT threat trends and the development of proactive defense strategies \cite{CtiView, Irshad2023}.

Building on these advancements, our unique approach leverages LLMs for CTI extraction and employs techniques such as zero-shot learning, chain-of-thought, and polymorphic prompting. This strategy allows for a one-time querying process, significantly enhancing resource utilization and speeding up the extraction of critical CTI elements. Unlike traditional methods, which often rely heavily on manual efforts and provide incomplete threat landscapes, our method offers clear, rationale-driven outputs that improve extraction accuracy and provide deeper insights for analysts. By incorporating these innovations, we deliver a more sophisticated and effective solution for navigating and addressing the complexities of the threat landscape.

\subsection{Ransomware Prevention and Mitigation}
Numerous studies have focused on identifying and mitigating ransomware threats \cite{McIntosh:2023,Oujezsky:2023,Thomas:2018,Chayal:2022, Kim2022}. Recognizing its effectiveness in safeguarding against this pervasive threat, the primary goal of these measures is to prevent and mitigate potential risks and security threats, effectively protecting systems, data, and networks from unauthorized access, breaches, or malicious activities \cite{Aldaraani:2018}.

Levesque et al. \cite{Levesque2014} developed a user-risk prediction model to identify potential malware targets by analyzing social, demographic, and behavioral factors, identifying twelve key features as crucial predictors. Similarly, Yilmaz et al. \cite{Yilmaz2023} explored the human element by investigating the relationship between personality types and ransomware victimization, though they found no significant correlation.

In contrast to these individual-focused studies, our solution emphasizes the broader context of company profiles and the operational characteristics of ransomware groups using the SKRAM (Skills, Knowledge, Resources, Authorities, and Motivation) model. This shift enables a more strategic approach to cybersecurity, allowing organizations to proactively adjust their defenses by understanding both the potential targets and the attackers. Additionally, our model generates a risk score for each identified threat, providing a quantifiable measure that organizations can use to prioritize and filter threats based on their potential impact. This holistic view significantly improves organizational resilience to ransomware attacks by reducing the noise in threat detection and focusing on the most pertinent threats.

\section{Methodology for Dataset Construction and Management}
\label{sec:datamanagement}

\subsection{In-Depth Analysis of Ransomware Campaigns}
Data collection for our study began with the aggregation of over 10,000 documented ransomware victim cases from Ransomware.live, a platform that compiles data from various open-source projects monitoring ransomware incidents since January 12, 2020 \cite{Ransomware:Live}. These records provide the ransomware group, the names of the victim company, and the dates of the attacks. Using this dataset, we initiated a filtering process to refine the selection of incidents for predictive modeling. Initially, we excluded all cases that occurred before 2021. We then omitted records related to threat actors not covered in our current dataset and discarded entries lacking detailed descriptions of the victim organizations. After applying these filters, we identified 409 distinct attacks for our predictive model.

The data features we extracted and summarized in Table \ref{tab:ml_data_features} were obtained through a dual approach: directly from victim records and supplemented by open-source databases. This method furnishes a profile for each victim, highlighting key attributes such as employee numbers and annual revenue. These details are critical as adversaries often leverage this information to gauge a company's capacity to meet ransom demands. For simplicity and practicality in organizational data management, we initially focused on static organizational data features. Including dynamic data features, which could offer a more in-depth understanding of a target's vulnerability to ransomware attacks, represents a potential area for future enhancement of our model. Adversary attributes, by contrast, present a more complex challenge for data extraction, requiring the sophisticated techniques detailed in the subsequent section to accurately capture these characteristics.

\begin{table}[htbp]
  \captionsetup{
    belowskip=8pt,
  }
  \rowcolors{2}{gray!7}{white} 
  \sffamily\small
  \centering
  \renewcommand{\arraystretch}{1.5}
  \begin{tabularx}{\textwidth}{|>{\hsize=.5\hsize}X|>{\hsize=1\hsize}X|>{\hsize=1.5\hsize}X|>{\hsize=1\hsize}X|}
    \hline
    \rowcolor{white}
    \textbf{Category} & \textbf{Data Feature} & \textbf{Description} & \textbf{Values} \\
    \hline
    Victim & Country Of Origin & The nation in which a product or service was manufactured, produced, or originated & ISO-3166 \\
    Victim & Industry Sectors* & Standard grouping used for economic analysis & STIX 2.1 industry sectors \\
    Victim & Annual Revenue & Monetary income within a fiscal year & Integer values \\
    Victim & Number of employees & Count of those employed by the organization & Integer values \\
    Victim & Org. type & Broader category such as for-profit, non-profit, school, hospital, etc. & Categorical values \\
    Adversary & Sophistication & Estimated degree of expertise in the tool and group & Categorical values \\
    Adversary & Motive & Broad objective, which is mostly financial gain & Categorical value \\
    Adversary & Intent & Tactical objective for damage to the victim & Categorical values \\
    Adversary & Resource level & Estimated level of people, money, and exploits behind the attacks & Categorical value \\
    Adversary & Capability & Count of MITRE ATT\&CK TTPs in use & Integer value \\
    \hline
  \end{tabularx}
  \caption[Summary of Data Features]{Summary of the data features used to build the machine learning model for ransomware victim prediction.}
  \label{tab:ml_data_features}
\end{table}

Adversary attributes were extracted from threat reports using our feature extraction system, with a focus on the STIX 2.1 Threat Actor Model attributes and SKRAM attributes. The utilization of the SKRAM model allows for a comprehensive understanding of threat actor capabilities and intentions, thereby providing valuable insights into their operational profiles. This approach not only enables a more nuanced analysis of the adversaries' skills, knowledge, available resources, authorities, and motivations but also supports the development of targeted defense mechanisms and proactive security strategies.

Our analysis included analyzing their motives (ranging from financial to ideological), capabilities and sophistication levels (to gauge their technical skills), and the resources at their disposal (spanning from individual actors to well-funded organizations). To deepen our understanding of these adversaries, we added attributes such as intent and quantified their known capabilities using the MITRE ATT\&CK framework. 

One feature which requires augmentation includes the level of activity of the ransomware group at the time of attack. Evaluating their objectives and past activities, it is clear ransomware groups launch campaigns across a timespan. Durations of no attacks typically indicate the group's functional status or possible disruption by law enforcement. This implies that even if an organization matches an adversary's victim profile, the adversary might not be actively attacking at that time. To capture this data feature, we use an exponential weighted moving average (EWMA). EWMA is a statistical method that helps in smoothing time series data, giving more weight to the most recent observations. The EWMA is calculated using the formula:
\begin{equation}
V_t = \lambda V_{t-1} + (1-\lambda) x_t
\end{equation}

Where $V_t$ is the moving average of attacks per month by an ransomware group at month $t$, which includes the weight $\lambda$ for the previous months EWMA and $1-\lambda$ for the current month. In our tests, we set $\lambda$ to 0.2 to simulate rapid decline after no observations in a given month. In both training and deployment, the EWMA is calculated over the entire dataset for each record. This approach ensures uniform measurement and provides a reliable basis for comparing activity levels across different time periods.

\subsection{Extraction of Ransomware Profiles Leveraging LLM Techniques}

To address the challenge of structuring narrative-based open-source intelligence on ransomware adversaries, we developed a methodology using LLMs to transform diverse data sources, including news articles and comprehensive threat analyses, into a machine-learning-friendly STIX format. Figure \ref{fig:llm_management} illustrates our approach involving a flexible prompt design strategy that interfaces with multiple LLMs. The responses from the LLMs are passed through the response processor schema for refining the raw output into formats ready for further analysis and eventual transformation into STIX-compliant records. This method enhances the adaptability and efficiency of our system, streamlining the development of new content processors and .ensuring the relevance and precision of the data in security applications.


For our study, we analyzed approximately 229 threat reports to identify SKRAM attributes for 146 distinct adversaries. While the actual number of unique adversaries may be lower due to aliases and overlapping identifiers, feature extraction was performed using OpenAI's \texttt{gpt-4-1106-preview} model, which has a context window size of 128,000 tokens and an output token limit of 4,096 tokens. These specifications enabled the model to process extensive contextual information from multiple reports or large segments of text in a single input, while still generating detailed and comprehensive outputs. To ensure the accuracy of the SKRAM features extracted by the LLMs, we engaged cybersecurity research students, who were compensated for their efforts, to review and verify the accuracy of the LLM-generated data.

\subsection{Designing the Prompt}
The prompting architecture is developed manually as reusable specifications designed to enhance learning accuracy through various approaches. It includes techniques to boost the reliability and accuracy of the LLM when extracting data features. Each feature extraction prompt is structured to include the key name of the feature, its extraction intent, and precise guidance, which are developed through the observation of LLM outputs as shown in the sample \ref{lst:feature-spec}. Additionally, the specification incorporates a set of examples following a many-shot prompt approach, improving output formatting. The process field outlines a sequence of logical steps that emulate a chain-of-thought reasoning pattern, aiding the LLM in feature extraction, as supported by Wei et al. \cite{wei2022chain}. Lastly, each feature extraction should align with a predefined standard, ensuring the extracted data is consistently formatted and immediately ready for integration into machine learning datasets.

\begin{lstlisting}[language=yaml, caption={Excerpts from a Feature Prompt}, label=lst:feature-spec, basicstyle=\scriptsize\ttfamily]
# Example target industry feature extraction under feature_keys
name: target_industry_sectors
intent: Identify the specific industry sectors...
guidance: Consider adversary objectives to determine industries...
examples:
  - sample: ...reveals a consistent focus on the financial sector...
    answer: [financial-services]
  - sample: ...particularly active against healthcare institutions...
    answer: [healthcare]
  - sample: Cyber espionage ... directed at technology companies...
    answer: [technology]
process:
  - ...understand which industries are primarily affected...
  - ...Assess the nature of the attacks and the objectives...
standard: Enumerated STIX Industry Sectors
\end{lstlisting}

Consistency in the JSON output is paramount for its application in machine learning. We achieve this by either referencing well-known open standards that are familiar to the LLM or by specifying a list of possible responses. For instance, we use the standard ISO 3166-1 alpha-2 codes for country identification and the MITRE ATT\&CK framework for defining threat attributes, mitigations, and potential impacts \cite{mitre2023attack}. Although there is no universally accepted standard for industry sectors, the STIX vocabulary provides a well-defined set of industry sectors that can be enumerated \cite{Stix:2.1}.

Moreover, our prompts are augmented to elicit a rationale for each feature extracted, mirroring the stop-and-think approach advocated in recent studies \cite{kojima2023large}.

\subsection{Processing LLM Response} \label{processing-llm-response}
\begin{figure}
  \captionsetup{
    belowskip=-8pt,
  }
  \centering
  \fbox{\includegraphics[width=0.985\linewidth]{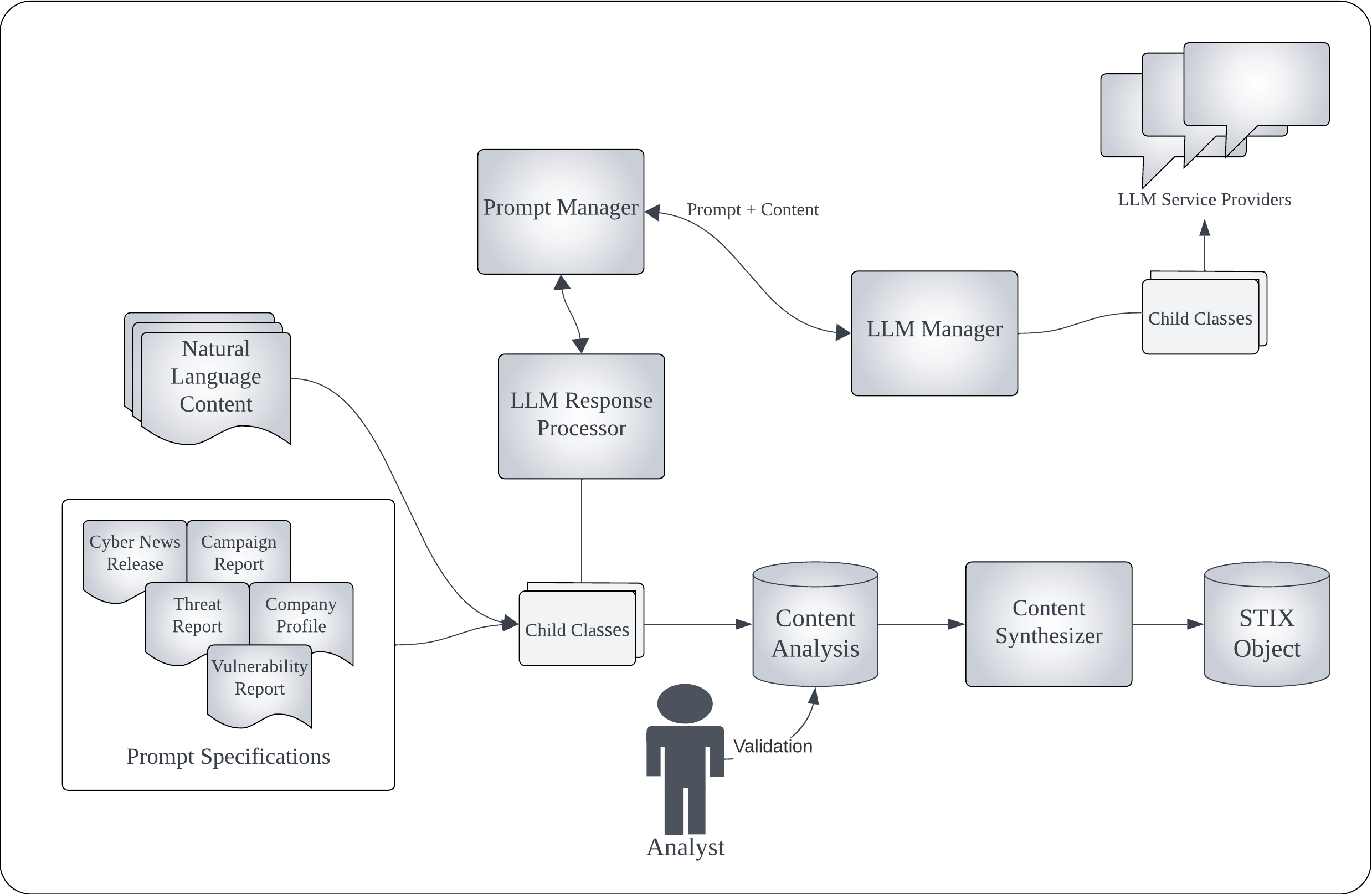}}%
    \caption{Class diagram illustrating the integration of prompt design and natural language content in our Chat Completion Feature Extraction System (CCFE).}
  \label{fig:llm_management}
\end{figure}

Due to the intricacies of our prompts, LLMs often generate multiple responses because of the limitations of output tokens. To address this, our LLM Response Processor, illustrated in Figure \ref{fig:llm_management}, serializes these responses to create a unified output, which improves coherence and consistency in the generated information. Furthermore, the LLM Processor ensures that the LLM response adheres to proper standards and validates enumerated standards, such as target sectors, as specified by our prompts. Any invalid or erroneous values are filtered out during processing. Additionally, vulnerabilities, adversary techniques, and mitigation identifiers are validated with the National Vulnerability Database (NVD) and Mitre Att\&ck to ensure their existence, thereby minimizing the propagation of misinformation to the analyst \cite{mitre2023attack, nvd}.

\subsection{Content Analysis}
 While fully automated feature extraction using LLMs is desirable, their non-deterministic nature can introduce potential inaccuracies. To address this, our analysts validate the SKRAM output to identify any inaccuracies, incompleteness, or outdated information \cite{Schlette2021}. They ensure that the output generated by the LLM accurately represents the content of the original report. This validation process significantly expedites the overall workflow, as analysts can review and confirm the extracted content directly. The Processing LLM Response module undertakes the bulk of this validation work, ensuring the removal of erroneous information and verifying that the extracted standards align with the expected criteria (see \ref{processing-llm-response} for details).

\subsection{Content Synthesizer}
Finally, the Content Synthesizer is specifically tailored for each type of prompt. This customization is critical for converting features in CTI reports into valid STIX objects. It provides a method to handle the diverse nature of CTI content effectively.

CTI reports, such as threat reports, contain information about vulnerabilities, targeted entities, and capabilities. These elements are accurately transformed into corresponding STIX objects. Similarly, campaign reports, which include details on tools, malware, and attack patterns, are processed into STIX objects and relationships.

The Content Synthesizer ensures that all CTI content is structured appropriately into STIX objects. These objects are then inserted into the STIX Database, facilitating their application and analysis.

\section{Dataset Augmentation via Synthetic Data Generation}\label{synth-date-generation}

The initial dataset's scope, limited to 409 ransomware victims, proved insufficient for accurate machine learning predictions. We addressed this through synthetic data generation, enhancing the dataset's volume while preserving the integrity of real-world insights.

First, to generate synthetic samples that closely resemble the characteristics of actual victims, each attack record replicates with the following process:
\begin{itemize}
    \item \textbf{Preserve Key Attributes:} For each victim, retain their original \texttt{country of origin}, \texttt{sectors}, and \texttt{company type} as these features are critical identifiers that should remain constant to maintain the integrity and relevance of the synthetic data.
    \item \textbf{Permute Numeric Attributes with Gaussian Noise:} Introduce variability into the \texttt{revenue}, \texttt{number of employees}, and \texttt{EWMA} attributes by adding Gaussian noise.
\end{itemize}

In our methodology, Gaussian noise was introduced to simulate the natural fluctuations in key numerical features, such as revenue and employee count, observed in real-world data. This noise was calculated based on the mean values derived from the actual data, ensuring that the distribution was centered around zero. The standard deviation was chosen to mirror the variability observed within the dataset, thereby enhancing the representativeness of the data while maintaining similarity to the characteristics of observed victims. By employing Gaussian noise in this manner, we successfully augmented our dataset to include ten times as many samples as the original, each enriched with diverse but plausible values, which broadened the scope of our analysis without deviating from the realism inherent to the victim data. The approach is presented in the following pseudo-code:

\begin{algorithm}
\caption{Generation of Synthetic Victims}
\begin{algorithmic}[1] 
\Procedure{GenerateSyntheticVictims}{$\mathbf{V}$, $\mathbf{R}$, $n$}
    \State $\mathbf{S} \gets \text{empty list}$ \Comment{$\mathbf{S}$ for synthetic victims}
    \ForAll{$v \in \mathbf{V}$} \Comment{Iterate over all victims in set $\mathbf{V}$}
        \For{$i \gets 1$ \textbf{to} $n$} \Comment{Generate $n$ synthetic profiles per victim}
            \State $r \gets \mathbf{R}[v.\text{'threat\_actor'}]$ \Comment{Select ransomware group $r$ for victim $v$}
            \State $s \gets \text{copy of } v$ \Comment{Create synthetic profile $s$ based on $v$}
            \State \Comment{Add Gaussian noise to attributes}
            \State $s.\text{'revenue'} \gets v.\text{'revenue'} + \mathcal{N}(\mu, \sigma^2)$
            \State $s.\text{'employees'} \gets v.\text{'employees'} + \mathcal{N}(\mu, \sigma^2)$
            \State $s.\text{'ewma'} \gets v.\text{'ewma'} + \mathcal{N}(\mu, \sigma^2)$
            \State $s.\text{'safe'} \gets 0$ \Comment{Mark synthetic profile $s$ as not safe}
            \State \Call{Append}{$\mathbf{S}$, $s$}
        \EndFor
    \EndFor
    \State \textbf{return} $\mathbf{S}$ \Comment{Return list of synthetic victims}
\EndProcedure
\end{algorithmic}
\end{algorithm}

To ensure a balanced model, we incorporated 'safe' samples, signifying entities that, with a high degree of certainty, are unlikely to be targeted by specific ransomware groups. These safe samples might represent organizations located outside the geographical or sectoral focus of the ransomware or entities that are potentially immune to an inactive ransomware variant, indicated by an EWMA approaching zero. The method involves generating an equivalent number of safe samples for each identified ransomware variant. This is achieved by introducing probabilistic variations to critical attributes, thereby crafting simulated profiles that reflect characteristics of non-targeted entities as follows:

\begin{algorithm}
\caption{Generation of Safe Synthetic Samples}
\begin{algorithmic}[1] 
\Procedure{GenerateSafeSamples}{$\mathbf{R}$, $country\_pool$, $company\_types\_pool$, $n$}
    \State $\mathbf{SafeSamples} \gets \text{empty list}$ \Comment{List to store safe samples}
    \ForAll{$variant, metrics \in \mathbf{R}$} \Comment{Iterate over ransomware variants}
        \ForAll{$victim \in metrics.victims$} \Comment{Iterate over each victim}
            \For{$i \gets 1$ \textbf{to} $n$} \Comment{Generate multiple safe samples}
                \State $sample \gets \text{copy of } victim$ \Comment{Create a base sample}
                \Repeat
                    \State \Comment{With probability $P(feature)$, modify the value to one without a victim}
                    \State $sample.\text{'origin'} \gets \Call{PermuteFeature}$
                    \State $sample.\text{'company\_type'} \gets \Call{PermuteFeature}$
                    \State $sample.\text{'revenue'} \gets \Call{AdjustNumericFeature}$
                    \State $sample.\text{'employees'} \gets \Call{AdjustNumericFeature}$
                    \State $sample.\text{'ewma'} \gets 0$ \Comment{With $P(ewma)$}
                    \State $sample.\text{'safe'} \gets 1$ \Comment{Mark the sample as safe}
                \Until{\Call{DidPermutate}{sample}}
                \State \Call{Append}{$\mathbf{SafeSamples}$, $sample$}
            \EndFor
        \EndFor
    \EndFor
    \State \textbf{return} $\mathbf{SafeSamples}$
\EndProcedure

\Function{DidPermutate}{$sample$, $metrics$}
    \For{$key \in \text{features to check}$}
        \If{$sample[key] \neq metrics.\text{original } key \text{ value}$}
            \State \textbf{return} $\text{True}$
        \EndIf
    \EndFor
    \State \textbf{return} $\text{False}$
\EndFunction

\end{algorithmic}
\end{algorithm}

In this algorithm, we assign probability weights to data features as seen in Figure \ref{tab:feature_weights}. These weights dictate the likelihood of each feature being altered to simulate a safe sample. With the probability weight of the feature, we will make the feature safe. Otherwise, we do nothing to the feature for the victim record. For example, the permutation for categorical features such as country of origin includes randomly sampling a country not present in the ransomware variant's victim dataset. Finally, we ensure at least one field has permuted in order to declare the record safe from any ransomware attacks.

\begin{figure}[ht]
\centering
\[
\begin{array}{l}
\text{{'country\_of\_origin': }} 0.8, \\
\text{{'number\_of\_employees': }} 0.3, \\
\text{{'revenue': }} 0.7, \\
\text{{'company\_type': }} 0.3, \\
\text{{'ewma': }} 0.95
\end{array}
\]
\caption[Dataset Feature Weights]{This table displays the weighting assigned to each feature used in the predictive model for ransomware victim likelihood.}
  \label{tab:feature_weights}
\end{figure}

Through this synthesis of real and synthetic data, we crafted a comprehensive and balanced dataset that significantly enhances the predictive model's reliability. This enriched dataset not only mirrors the complexity and variability of real-world data but also significantly bolsters the predictive accuracy of our model. 

\section{Ransomware Risk Machine Learning}
\label{sec:analysis}
In this section, we detail the development and results of a machine-learning model designed to predict an entity's ransomware risk. To deploy the model, we opted for a Random Forest Classifier due to its performance in tackling classification tasks to discern between safe and unsafe entities, which is then translated into a risk score. Random Forests stand out for their resilience to overfitting, attributed to their ensemble-based methodology, which mitigates the risk of memorizing training data patterns and facilitates better generalization to unseen data. Moreover, Random Forests demonstrate proficiency in handling high-dimensional datasets containing numerous features. Their adeptness in managing complex data structures ensures our model's capability to capture subtle relationships between the various safe and unsafe entities.

\subsection{Model Configuration and Training}
The Random Forest model was configured with n\_estimators set to 100 and a random\_state of 42. We used 100 trees to balance accuracy with computational efficiency, considering the size of our dataset. The specific random\_state ensures consistent results across multiple runs of the model. The dataset, consisting of 8,200 records evenly split between safe and unsafe instances, was divided using an 80/20 split for training and testing.

Categorical columns with multiple labels were binarized, and single-label string categorical columns were encoded using one-hot encoding. Non-categorical features were label-encoded to ensure uniform data formatting conducive to effective model input. It is important to note that as new industry sectors and countries are introduced into our dataset, the model will undergo periodic retraining. This adaptation helps maintain its accuracy in assessing the likelihood of an entity being targeted by a ransomware group.

\subsection{Results}

The resulting model achieves high precision, recall, and F1-score, all measuring at 99\%. In analyzing the confusion matrix, we observe the model correctly predicted 807 instances as negative and 820 instances as positive. It incorrectly classified 13 instances as positive but did not mislabel any instances as negative \ref{tab:confusion_matrix}.

\begin{table}[htbp]
  \centering
  \renewcommand{\arraystretch}{2} 
  \begin{tabular}{cc|c|c|}
    \cline{3-4}
    & & \multicolumn{2}{c|}{\textbf{Predicted}} \\ \cline{3-4} 
    & & \textbf{Negative} & \textbf{Positive} \\ \hline
    \multicolumn{1}{|c|}{\multirow{2}{*}{\textbf{Actual}}} & \textbf{Negative} & \hspace{8pt}807\hspace{8pt} & \hspace{8pt}0\hspace{8pt} \\ \cline{2-4} 
    \multicolumn{1}{|c|}{} & \textbf{Positive} & \hspace{8pt}13\hspace{8pt} & \hspace{8pt}820\hspace{8pt} \\ \hline
  \end{tabular}
    \vspace{10pt} 
    \caption{Random Forest Confusion Matrix: 807 instances were correctly predicted as negative, 820 instances were correctly predicted as positive, 13 instances were falsely predicted as positive, and there were no instances falsely predicted as negative.}
  \label{tab:confusion_matrix}
\end{table}

The feature importance depicted in Figure \ref{fig:feature_importance} highlights the entity attributes that significantly influence the risk score. The EWMA, although removed from the chart to underscore the importance of the company profile, was the most significant feature, as seen by the predictions in Figure \ref{fig:safe_victim_output} and Figure \ref{fig:company_victim_output}. The likelihood of a ransomware group targeting an entity decreases substantially if the group is inactive. This is followed in significance by the company's location, size, revenue, and industry sectors. Despite this, the chart shows that the company profile can override the EWMA feature if the company’s size, location, and industry do not align with the ransomware group's profile. However, the prominence of US-based companies in the feature importance indicates a potential internal bias in our dataset, suggesting a possible imbalance in the distribution by country of origin.

\subsection{Ransomware Prediction}\label{ransomware_predictions}
\begin{figure}
  \captionsetup{
    belowskip=-8pt,
  }
  \centering
  \includegraphics[width=0.9\textwidth]{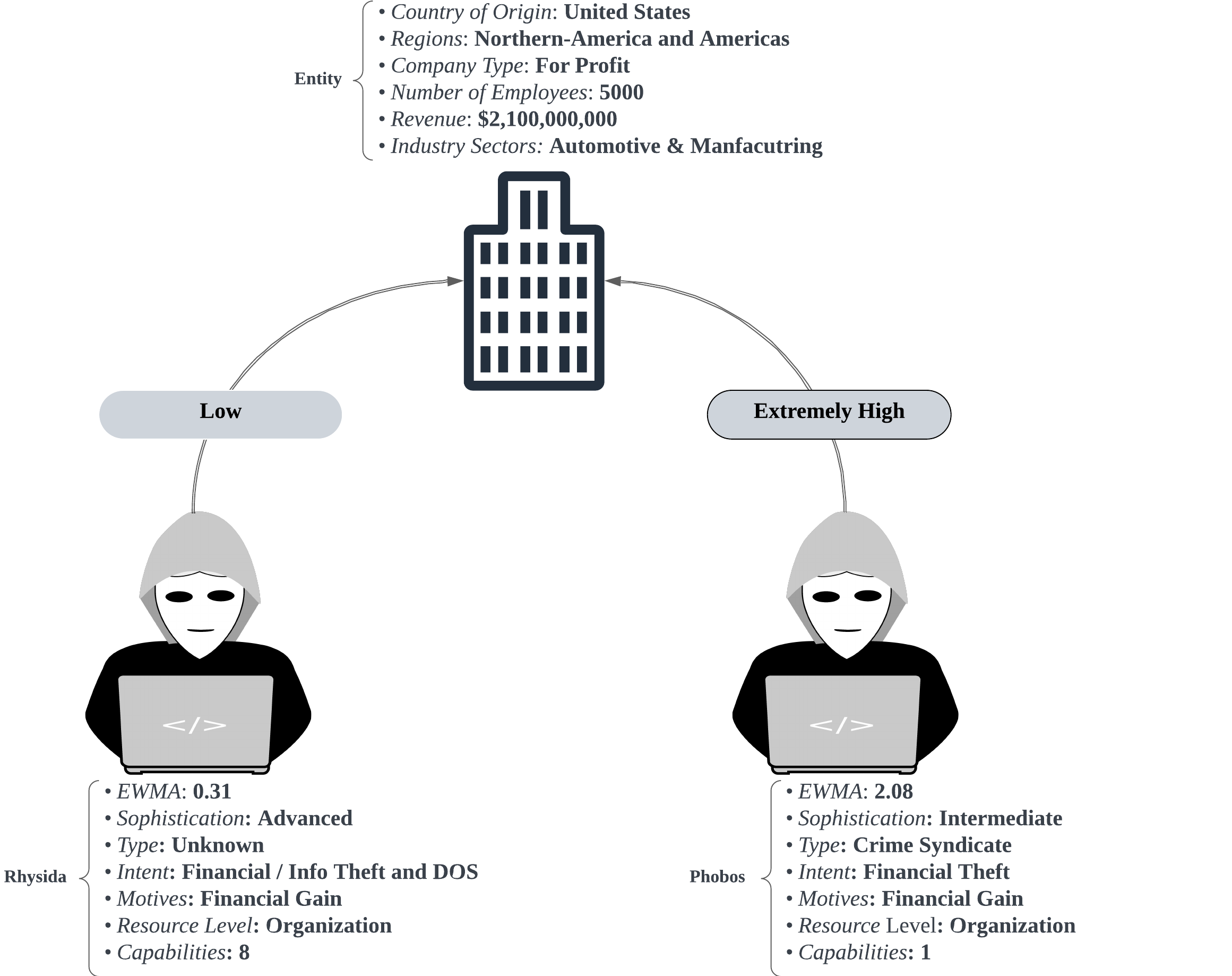}%
\caption[Ransomware Victim Prediction]{This image illustrates the predictions for the Ransomware Groups Phobos and Rhysida regarding their likelihood of targeting an entity. The feature importance for the prediction with an "Extremely High" likelihood is detailed in Figure \ref{fig:company_victim_output}, while the prediction with a "Low" likelihood is detailed in Figure \ref{fig:safe_victim_output}.}
  \label{fig:ransomware-prediction}
\end{figure}

\vspace{6pt}
This section explores two predictions generated by our Ransomware Prediction Model. The model produces a confidence score that assesses the likelihood of a company being a potential target based on how closely their profile aligns with those typically targeted by a specific ransomware group. This assessment also considers the current level of activity of the ransomware group. The confidence score is then converted into a 10-point risk scale, ranging from None (0) to Extremely High (9). The data presented in Figure \ref{fig:ransomware-prediction} depict predictions for the same U.S.-based company, for two different ransomware groups. This company operates within the Automotive and Manufacturing sectors, employs 5,000 people, and generates an annual revenue of \$2,100,000,000.

\begin{figure}
  \captionsetup{
    belowskip=-8pt,
  }
  \centering
  \includegraphics[width=\textwidth]{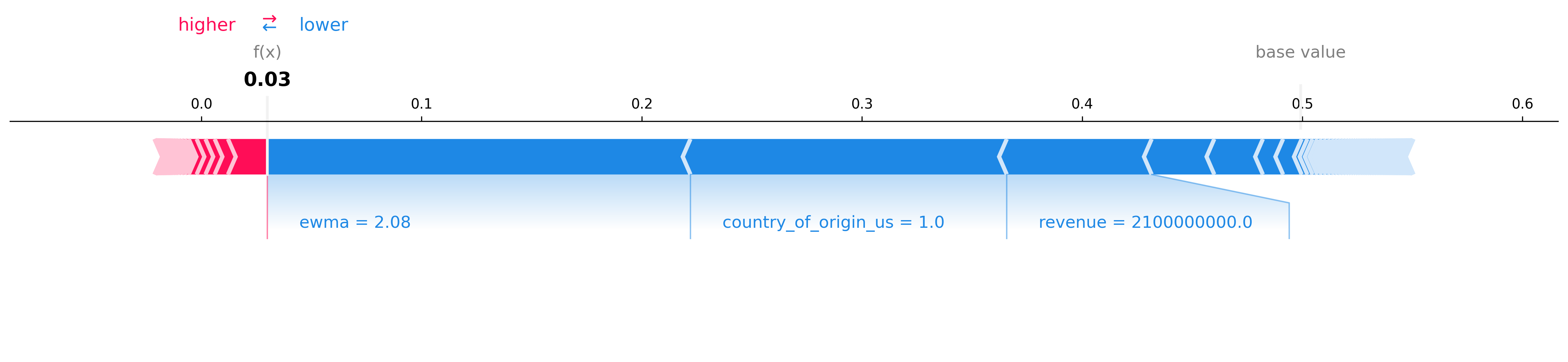}%
\caption[Extremely High Prediction]{
The chart illustrates the feature importance of the entity detailed in Figure \ref{fig:ransomware-prediction} who has an 'Extremely High' likelihood of being targeted by the Ransomware Group Phobos.}
  \label{fig:company_victim_output}
\end{figure}

The 'Extremely High' prediction presented in Figure \ref{fig:ransomware-prediction} provides a prediction concerning the ransomware group Phobos, which is identified as an intermediate-level "crime syndicate." The group's EWMA score is 2.08 indicates active participation in malicious activities. Notably, this threat actor possesses a singular capability, primarily aimed at financial gain and information theft, which marks it as a well-resourced entity. As illustrated in Figure \ref{fig:company_victim_output}, the extremely high risk score assigned to this record stems from the threat actor's on-going activity and strategic focus on targeting for-profit entities in the United States, especially those comparable in size regarding employee numbers and annual revenue.

\begin{figure}
  \captionsetup{
    belowskip=-8pt,
  }
  \centering
  \includegraphics[width=\textwidth]{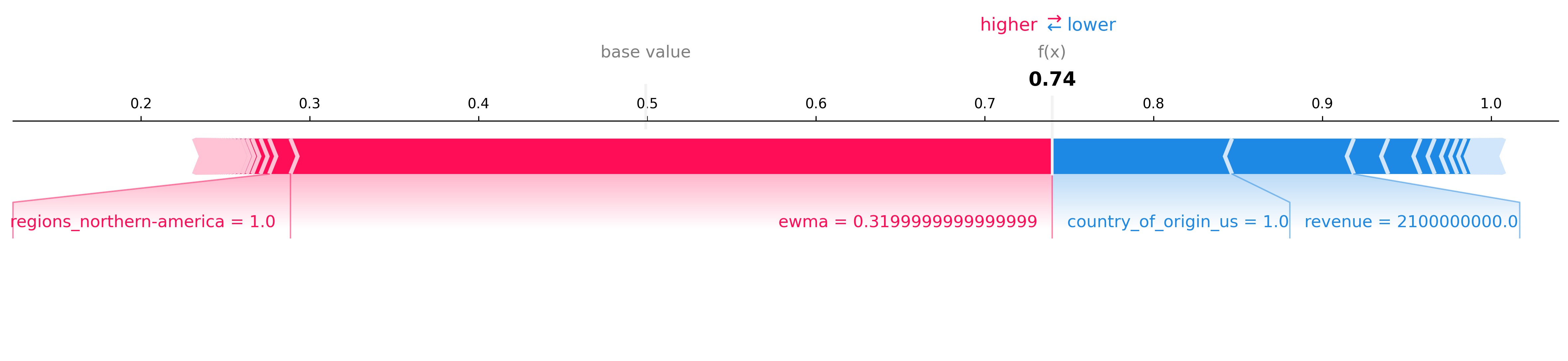}%
\caption[Low Prediction]{
The chart illustrates the feature importance of the entity detailed in Figure \ref{fig:ransomware-prediction} who has a 'Low' likelihood of being targeted by the Ransomware Group Rhysida.}  
\label{fig:safe_victim_output}
\end{figure}

However, the 'Low' prediction shown in Figure \ref{fig:ransomware-prediction} details a prediction for the ransomware group Rhysida. This group is characterized as advanced, with motives including financial theft, disruption of service, and information theft, all driven by financial gain. The group is well-resourced, possessing eight known capabilities. While the group does show activity, the prediction received a "Low" risk rating primarily due to the profile of this entity. As seen in Figure \ref{fig:safe_victim_output}, the entity in question does not fit the typical victim profile of the ransomware group, which usually does not target companies of this size within the United States, specifically those in the automotive sector.

\subsection{Dataset Considerations}

In reflecting upon our model's high accuracy and its implications for generalizability, it is imperative to consider the role of synthetic data in our study. While synthetic data has significantly expanded our training dataset, enabling a comprehensive analysis that would not be possible with the limited number of real cases alone, this approach inherently introduces questions about the representativeness of our findings. The synthetic generation process aims to mimic the diversity of real-world scenarios, yet the extent to which it captures the nuanced behaviors of ransomware attackers and victim entities remains an area for further scrutiny. To address this, future work will focus on validating model predictions with more extensive datasets of real incidents as they become available, enhancing our understanding of the model's applicability across different contexts.

\begin{figure}
  \captionsetup{
    belowskip=-8pt,
  }
  \centering
  \fbox{\includegraphics[width=0.985\linewidth]{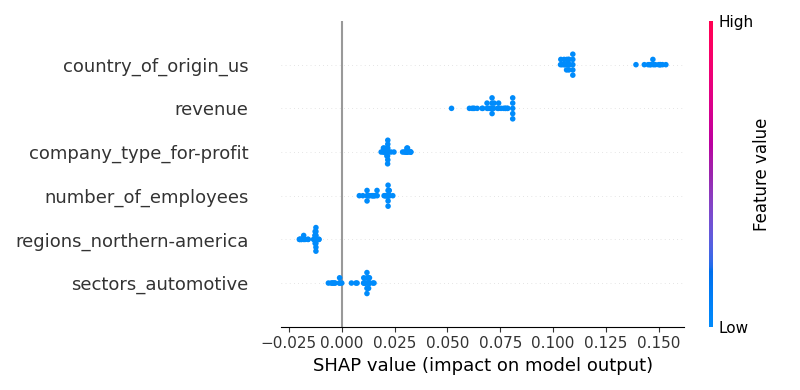}}%
  \caption[Ransomware Prediction Feature Importance]{This figure illustrates the top 6 features, deliberately excluding the Enumerated Weighted Moving Average (EWMA) of the Ransomware's Group. Underscoring the relationship between ransomware groups and their victims in predicting the likelihood of being targeted by a ransomware group. }
  \label{fig:feature_importance}
\end{figure}

Furthermore, the analysis, as illustrated in \ref{fig:feature_importance}, reveals a pronounced over-representation of US-based companies in our dataset. Importantly, this geographical bias is not a result of the original composition of data but stems from our data annotation process. Faced with incomplete or scarce information on any company, our protocol has been to exclude such entities to uphold the dataset's integrity and ensure analytical precision. While this decision is crucial for maintaining the quality of our analysis, it inadvertently introduces a geographical bias, potentially affecting our model's ability to accurately predict ransomware risks for organizations in less represented regions or countries.

\section{Future Work}
\label{sec:futurework}
In considering future research directions, a critical focus lies in automating, expanding, and diversifying the dataset used for analysis. Currently, our dataset consists of 8200 records, evenly split between 4100 safe and 4100 unsafe entities, all derived from 409 manually annotated authentic ransomware victims. There is a significant opportunity to enhance the dataset's size and diversity.

Future efforts should focus on transitioning away from reliance on synthetic data generation, moving towards automating the incorporation of real-world ransomware victim data. This transition would not only introduce greater diversity into the dataset but also enable a broader spectrum of risk analysis. Automating this process would further ensure that the ransomware group's EWMA and victim profiles remain current, reflecting the latest threat landscapes.

Additionally, exploring the performance of a growing number of LLMs such as Google's Gemini \cite{gemini}, Sanctuary AI's Grok \cite{grok}, Anthropic's Claude \cite{anthropic}, and Meta's Llama \cite{llama2}. A comparative analysis of these models would assess their effectiveness in capturing, processing, and serializing threat intelligence compared to OpenAI's gpt4.

Incorporating features to predict the potential financial impact and negotiation dynamics with ransomware groups could significantly enhance existing cybersecurity models. This approach would provide detailed assessments of the economic consequences and the likelihood of successful negotiations following an attack, offering a more comprehensive understanding of the threat landscape and aiding organizations in strategic decision-making and risk management.

\section{Conclusion}
\label{sec:conclusion}
In response to the growing threat of ransomware attacks, we propose the utilization of a Random Forest Classifier and LLMs to equip organizations with proactive risk assessments. Our proposed solution harnesses the capabilities of GPT-4 to efficiently distill and automate high-level strategic intelligence about ransomware groups. We pair this intelligence with data from Ransomware.Live, enabling the creation of profiles of the entities targeted by these ransomware groups. Our model provides entities with risk scales indicating their likelihood of being targeted, effectively filtering out inactive groups. However, it is crucial to undertake further ethical assessments before employing these findings to inform mitigation strategies. Inaccuracies may result from features extracted from LLMs, necessitating additional scrutiny to effectively validate and refine the outputs. For ethical treatment of the results, systems should acknowledge the role of AI in this context as a tool for risk prediction rather than a definitive source of truth. Decision-makers should consider AI-generated insights as one component in a broader decision-making framework for defensive actions and prioritization.


\section*{Acknowledgments}
This work was supported in part by the Department of Energy under award number \href{https://doi.org/10.13039/100000015}{DE-CR0000022}.

\printbibliography
\end{document}